\documentclass[fleqn,twoside]{article}
\usepackage{amsmath}  
\usepackage{amssymb} 
\usepackage{espcrc2}

\usepackage{graphicx}
\usepackage[figuresright]{rotating}

\newcommand{\be}{\begin{equation}}
\newcommand{\ee}{\end{equation}}
\newcommand{\bes}{\begin{equation*}}
\newcommand{\ees}{\end{equation*}}

\title{QCD at small baryon number} 

\author{Slavo Kratochvila\address[ETHZ]{Institute for Theoretical Physics, ETH Z\"{u}rich, CH-8093 Z\"{u}rich, Switzerland}\thanks{Talk presented by S. Kratochvila} and
        Philippe de
        Forcrand\addressmark[ETHZ]\address[CERN]{CERN, Physics Department, TH Unit, CH-1211 Geneva 23, Switzerland}}

\begin{document}

\markright{\hfill\rm CERN-PH-TH/2004-176\quad\quad}

\begin{abstract}

We consider the difficulties of finite density QCD from
the canonical formalism. We present results for
small baryon numbers, where the sign problem can be controlled, in
particular by supplementing the $\mu$=0 sampling with \linebreak imaginary $\mu$
ensembles. We initiate the thermodynamic study of few-nucleon
systems, starting with the measurement of the free energy of a few baryons
in the confined and deconfined phase.
We present a simple model for the phase transition, whose results are in good
agreement with the literature, but extend to lower temperatures.
\vspace*{-0.20cm}

\end{abstract}

\maketitle

\section{INTRODUCTION}

QCD at finite density is of great physical interest, but numerically challenging:
the usual Monte \linebreak Carlo sampling weight becomes complex, causing a sign problem.
Recently, progress has been made using different approaches \cite{Fodor:2001au,Allton:2002zi,deForcrand:2002ci,D'Elia:2002gd}.
However all these methods have severe limitations which are already being reached, with little prospect for further progress.

Therefore, we propose yet another method - in a canonical formalism.
Physically, we are interested in the study of systems consisting of a few
baryons,
for example the formation of nuclear matter or the binding of Deuteron.
Here, we measure the free energy of sectors with small baryon number, and present a simple model for the
QCD deconfinement phase transition.
Numerically, our sign problem is governed by the baryon number $B$,
and \emph{not} by the volume, and its dependence on temperature is mild.
The sector $B=0$ (no sign problem) was analysed in detail in \cite{Kratochvila:2003rp}.
Here, as a continuation, we present results for $B=1,2,\dots$
\vspace*{-0.25cm}

\section{QCD'S PARTITION FUNCTIONS}

The  grand canonical partition function with fixed chemical potential $\mu$
\be \label{eq:partitionfunctions_ZGC}
Z_{ GC }(\mu) = \int [DU] \; e^{-S_g[\beta,U]} \det M(U;\mu)
\ee
is numerically problematic for real chemical potential, since $\det M(U;\mu)$ becomes complex.
However, for imaginary $\mu=i \mu_I$, $\det M(U;i \mu_I)$ remains real.
Moreover, $Z_{ GC }(i \mu_I)$ is even and $\frac{2\pi T}{3}$-periodic
in $\mu_I$ \cite{Roberge:mm}.

The canonical partition function with fixed quark number $Q$ can be obtained from the grand canonical one using
a Fourier transform,
\vspace{-0.5cm}

\be
\label{eq:partitionfunctions_ZC}
Z_{ C}(Q) = \frac{1}{2\pi}\int_{-\pi}^{\pi} d \left( \frac{ \mu_I }{ T }\right)\;
e^{-i Q \frac{\mu_I}{T} } Z_{ GC }(\mu =  i \mu_I )\;.
\ee
\vspace{-0.5cm}

The $\frac{2\pi T}{3}$-periodicity of $Z_{GC}(i\mu_I)$ implies that $Z_{C}(Q  \neq 0 \mod 3 ) = 0$.
Hence, we only deal with integer baryon number ${B \equiv Q/3}$.

Knowing the canonical partition functions, $Z_{GC}(\mu)$ can then be calculated for arbitrary chemical potential
using the fugacity expansion (Laplace transformation):
\vspace{-0.5cm}

\be \label{eq:partitionfunctions_ZGC_statmech}
Z_{GC}(\mu) = \sum_{B=-\infty}^{\infty} e^{-3 B \frac{\mu}{T}} Z_C(B)\;.
\ee

\section{THE METHOD}

Our main observable is the free energy per baryon, $F(B)$, defined by $\frac{Z_C(B)}{Z_C(B=0)} \equiv e^{-\frac{1}{T}F(B)}$.
We obtain $Z_C(B)$ by following the approach in \cite{Hasenfratz:1991ax}, which samples $Z_{GC}(i\mu_I)$ at
fixed $\mu_I$'s, then Fourier transforms each $\det(U;i\mu_I)$ (all the $\mu_I$-dependency is in the determinant),
\vspace{-0.5cm}

\be \label{eq:loop_expansion}
    \det M(U;i\mu_I) = \sum_{B=-V}^V \hat Z_C(U;B) e^{-i 3 B \frac{\mu_I}{T}}.
\ee
\vspace{-0.5cm}

\noindent
Substituting into Eq.~(\ref{eq:partitionfunctions_ZGC}) then into
Eq.~(\ref{eq:partitionfunctions_ZC}) gives
\vspace{-0.5cm}

\bes
    Z_{C}(B) = \int [DU] \hat Z_C(U;B) e^{-S_g[U]}\;.
\ees
\vspace{-0.5cm}

We sample the grand-canonical ensemble at couplings $\beta$ and $i\mu_I$,
hence the observables we measure are ratios of partition functions,
\vspace{-0.25cm}
\bes
 \frac{Z_C(\beta,B)}{Z_{MC}(\beta,i\mu_I)} = \langle  \frac{\hat Z_C(B,U)}{\det M(U;i\mu_I)}   \rangle_{\beta,\mu_I}\;.
\ees
\vspace{-0.25cm}

In \cite{Crompton:2002wf}, the idea of the existence
of a particular $\mu_I(B)$ to extract $Z_C(\beta,B)$ optimally has been formulated.
Here, in order to improve the accuracy on the $Z_C(\beta,B)$'s, we simply combine results
from several $(\beta,\mu_I)$-ensembles by Ferrenberg-Swendsen reweighting\cite{Ferrenberg:yz}.
This requires knowing the determinant for different $\mu_I$'s.
This information comes as a by-product of our determination of
the $\hat Z_C(B)$ in Eq.~(\ref{eq:loop_expansion}).
In the temporal gauge ($U_4({\mathbf x},t)={\mathbf 1}$ except for $t=N_t-1$),
the staggered fermion matrix $M$ can be written as
\footnotesize
\vspace{-0.25cm}

\bes
  M = \left( \begin{matrix}
        B_0 & {\mathbf  1} & 0 & ... & 0 & U^\dagger_{N_t-1} e^{-a \mu N_t}\\
         - {\mathbf  1} & B_1 &  {\mathbf  1} & 0 & ... & 0\\
         0 & - {\mathbf  1} & B_2 &  {\mathbf  1} &  0& ... \\
          &  &  &   ... &  & \\
         -U_{N_t-1}e^{a \mu N_t} & 0 & ... & 0 &  - {\mathbf  1} &  B_{N_t-1}
      \end{matrix}
  \right) \normalsize
\ees
\normalsize
\vspace{-0.25cm}

Following \cite{Gibbs:1986hi}, we define a fermionic transition matrix
\vspace{-0.7cm}

\bes
 P = \left(
        \prod_{j=0}^{N_t-1} \left( \begin{matrix}
            B_j & {\mathbf  1} \\
            {\mathbf 1} & 0
            \end{matrix} \right)
       \right) U_{N_t-1}
\ees
\vspace{-0.25cm}

\noindent
with eigenvalues $\lambda_1, \ldots, \lambda_{6V}$.
We obtain the $\hat Z_C(U;B)$ using Eq.~(\ref{eq:loop_expansion}) and the equality
\begin{align}
    \det M(U;\mu) &= e^{3 V \frac{\mu}{T}} \prod_{i=1}^{6 V}\left( \lambda_i + e^{-\frac{\mu}{T}} \right) \;. \nonumber
\end{align}
\vspace{-0.95cm}

\section{RESULTS}
\vspace{-0.25cm}

Since this is a feasibility study, we consider a small ($4^4$)
lattice  with volume $V\sim(1.2~\text{fm})^3$.
We use four flavours of staggered quarks with mass $m = 0.05
a^{-1}$. This theory is expected to have
a first-order phase transition in the whole $(T,\mu)$ plane. With Hybrid
Monte Carlo,
we generate 55 ensembles of 4000--16000 configurations, at 11
temperatures $\frac{T}{T_c} \in [0.8, \ldots, 1.1] $, each at
five imaginary chemical potentials.

\begin{figure}[h!]\label{fig:results_freenrgperbaryon_beta}
\vspace{-0.6cm}

\begin{center}

\includegraphics[width=5.00cm,angle=-90]{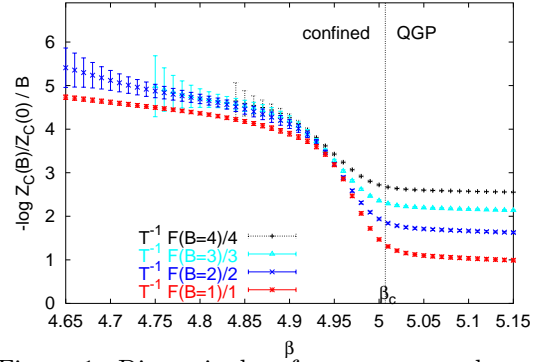}
\vspace{-1.0cm}

\caption{Dimensionless free energy per baryon, $T^{-1} F(B)/B$, as a function of $\beta=6/g^2$.}
\end{center}
\vspace{-1.00cm}

\end{figure}

In the high-temperature phase $T>T_c$, we observe the characteristics of
a system of free, relatively light quarks:
the free energy $F(\mu)$ is proportional to $\mu^2$ for small
$\mu$ \cite{Kratochvila:2003rp}, so that
the free energy $F(B) \sim B^2$ for small $B$ ($\mu$ is conjugate to
$B$).
Furthermore, $F(B)$ is almost independent of temperature in
this regime. In Fig.~1,
we show $T^{-1} F(B)/B$ as a function of $\beta$ for various $B$,
and indeed observe
almost equidistant, $\beta$-independent levels ($F(B)/B \sim B$).
For entropy reasons,
we expect $F(B=1)$ to vanish in the thermodynamic limit.
Deviations from these expectations are caused by finite size effects.

As $T$ approaches $T_c$ from below, the free energy drops dramatically.
This is caused by an increase in the number of thermally accessible
states, which triggers an entropic catastrophy \`a la Hagedorn.
This collective phenomenon (of gluons) is visible even in the
1-baryon sector.

In the low-temperature phase $T<T_c$, our results are consistent
with a simple model of weakly interacting heavy baryons at rest
(${p_{min} = \frac{2\pi}{4} a^{-1} \gg T = \frac{1}{4}a^{-1}}$).
At fixed temperature, $F(B)/B$ stays constant over a whole range
of $B \geq 2$, so that the free energy grows linearly with $|B|$
as sketched in Fig.~2. The interaction is repulsive: $F(B)/B -
F(B=1) \approx 0.1 a^{-1} \sim  60$ MeV per baryon, see Fig.~1.
In the presence of a (real) quark chemical potential $\mu$, the
free energy acquires an additional term $3 \mu B$, which tilts the
free energy in Fig.~2. The free energy minimum jumps away from
$B=0$, and a first-order transition to a finite density occurs,
when \vspace{-0.5cm}

\be \label{eq:criticalmu}
\mu_c =  \frac{F(B)}{3 B}|_{\text{plateau}}\;.
\ee

\begin{figure}[h]\label{fig:results_linearfreenrg}
\vspace{-2cm}

\begin{center}
 \hspace*{5.750cm}\includegraphics[height=5.7cm]{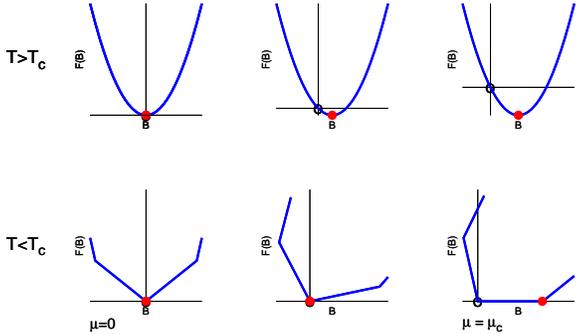}
 \vspace{-1.0cm}

\caption{Sketch of the free energy $F(B)$ in the two phases, for
increasing (real) chemical potential. For $T<T_c$, the free energy minimum jumps
to a non-zero density at a critical $\mu_c$.}
\end{center}
\vspace{-1.00cm}

\end{figure}

The phase diagram resulting from our simple model is shown in Fig.~3.
The agreement with other
methods for $\mu \lesssim 0.35 a^{-1}$ is very good, except at low
$\mu$, where our method suffers from finite size effects.
In the thermodynamic limit, $F(B=1)$ will develop a singularity
exactly at the $(T=T_c,\mu=0)$ transition.
Moreover, in the limit $T \to 0$, $F(B=1) = m_N$,
yielding via Eq.~(\ref{eq:criticalmu})
$\mu_c = m_N/3$, which already is a good approximation of the true phase
transition. Therefore, we expect Eq.~(\ref{eq:criticalmu})
to give a good estimate of the critical chemical potential in the
whole $(T,\mu)$ plane. On our small lattice, a plateau for $F(B)/B$
is already reached for $B=2$. The resulting phase boundary is shown
in Fig.~3. Note the bending down for $\mu/T > 1$, where the reliability
of other methods may be questioned. \\

%
%
%
%
\vspace{-0.75cm}

\section{CONCLUSION}
\vspace{-0.15cm}

In 4-flavor QCD, we have observed that the free energy grows linearly
with the baryon number, for small densities at $T<T_c$.
This behavior is consistent with a first-order transition from weakly
interacting heavy baryons to light, quasi-free quarks.
The corresponding phase boundary agrees with the literature for moderate
$\mu/T$, then bends downward.

The same information could, in principle, be obtained by reconstructing
the full grand-canonical partition function Eq.(\ref{eq:partitionfunctions_ZGC_statmech}). However, the sectors
with large baryon number, which we did not consider, bring in little
information but a lot of noise. This is why our approach works better.

Since the sign problem is moderate, we can increase the volume and lower
the temperature. As we do so, we expect the weak Pauli repulsion
among baryons to be overcome by nuclear attraction, leading to
the formation of nuclear matter.
%
%

\markright{}

\begin{figure}[t!]\label{fig:results_phasediag}
\vspace{-0.6cm}

\begin{center}

\hspace*{-0.5cm} \includegraphics[width=5.50cm,angle=-90]{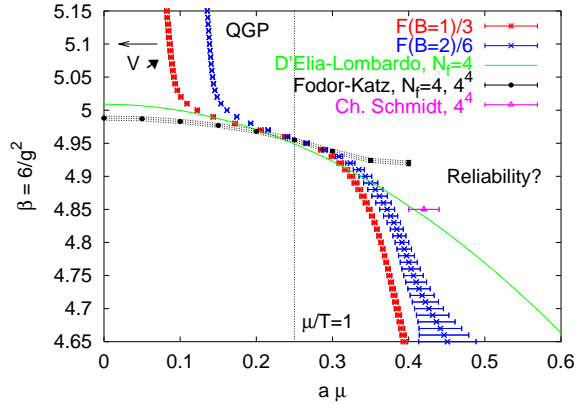}
 \vspace{-1.05cm}

\caption{\protect{The $(\mu,T)$ phase diagram.
An approximation of the phase boundary is given by $F(1)/3$,
where we neglect the baryonic interactions.
$F(2)/6$ already gives the plateau-value in
Eq.~(\ref{eq:criticalmu}).
Also shown are \cite{Fodor:2001au} and \cite{D'Elia:2002gd}, which use
the same simulation parameters.
Agreement is very good for $\mu \lesssim 0.35 a^{-1}$.
Ch.~Schmidt recently obtained one point\cite{Schmidt:2004ke}
using the factorization method\cite{Ambjorn:2002pz}.
}}
\end{center}
\vspace{-1.35cm}

\end{figure}

\end{document}